\documentstyle[psfig]{article}
\newcommand{\vs}{\vspace{-0.1cm}}
\newcommand{\vsss}{\vspace{-1cm}}
\begin{document}
\title{\vsss\vsss The Granular Phase Diagram\footnote{J. Stat. Phys. (in press)}}
\author{Sergei E. Esipov,\\
James Franck Institute and the 
Department of Physics, 
University\\ of Chicago, 
5640 S. Ellis avenue, Chicago, IL 60637, USA\\
Thorsten P\"oschel,\\
Humboldt-Universit\"at zu Berlin, 
Institut f\"ur Physik, \\
Invalidenstra\ss e 110,
D-10115 Berlin, Germany}
\date{August 23, 1996}
\maketitle
\setlength{\baselineskip}{18pt}
\begin{abstract}
\setlength{\baselineskip}{18pt}
  The kinetic energy distribution function satisfying the Boltzmann
  equation is 
  studied analytically and numerically
  for a system of inelastic hard spheres in the case of
  binary collisions. Analytically, this function is shown to have 
  a similarity form
  in the simple cases of uniform or steady-state flows. This determines
  the region of validity of hydrodynamic description. The latter
  is used to construct the phase diagram of granular systems,
  and discriminate between clustering instability and inelastic
  collapse. The molecular dynamics results support analytical results, 
  but also exhibit a novel fluctuational breakdown of
  mean-field descriptions. 
\end{abstract}
\thispagestyle{empty}
Granular media such as sand provide an attractive opportunity to
revisit a number of topics in classical physics, and contribute
new angles. In this paper we describe the granular phase diagram
which we hope will be helpful to a broad community given the
raising interest in granular systems ~\cite{Behringer}. 
The researchers interested in 
diluted granular gases, such as in astrophysical
applications ~\cite{astro}, and researchers who study, say, 
compaction of sand ~\cite{Knight} use different approaches.
The phase diagram may represent a ground for communication.

\begin{figure}[p]
\centerline{\psfig{figure=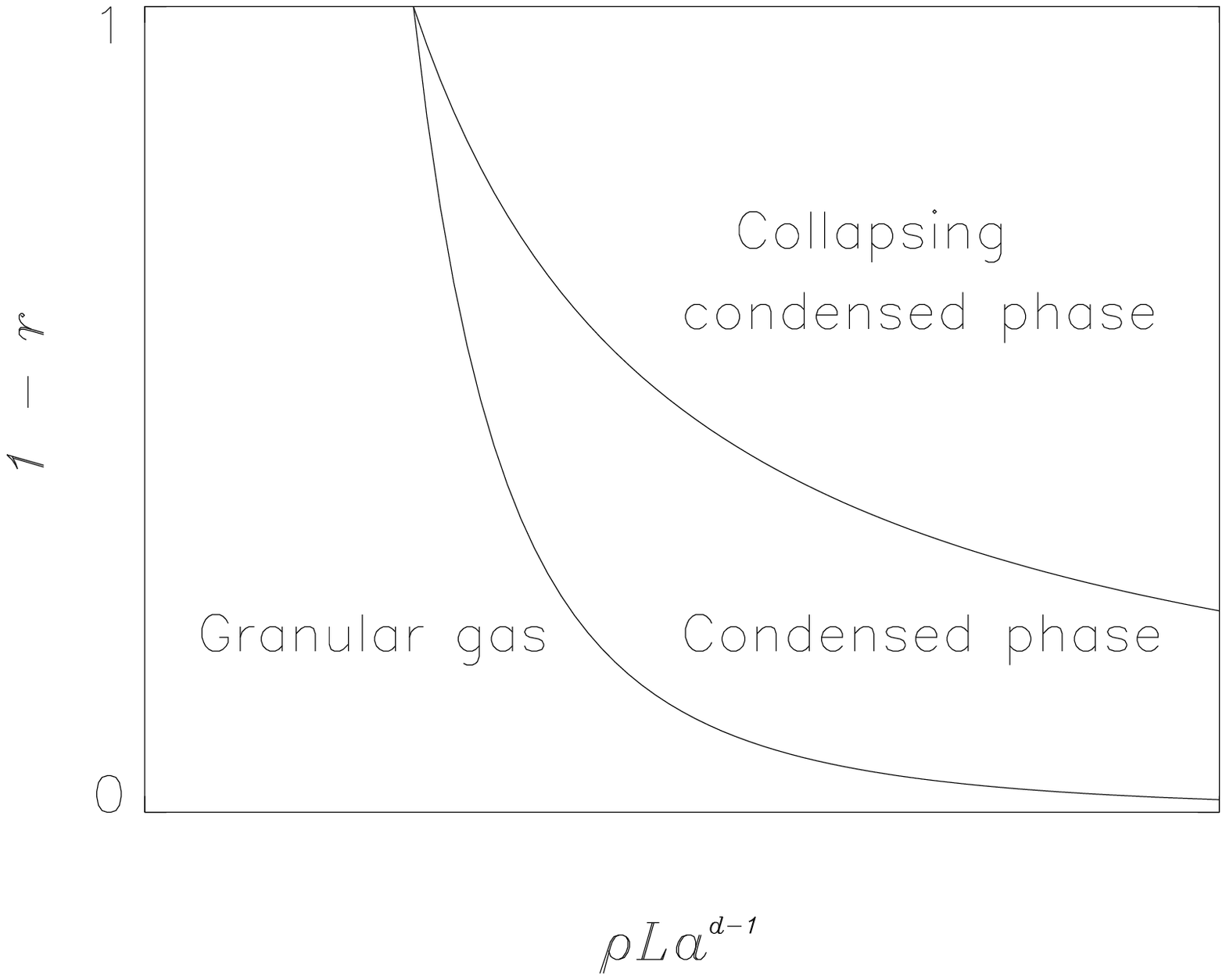,width=14cm,angle=90}}
\caption{The granular phase diagram. See text for details.}
\label{fig1}
\end{figure}

\begin{figure}[p]
  \centerline{\psfig{figure=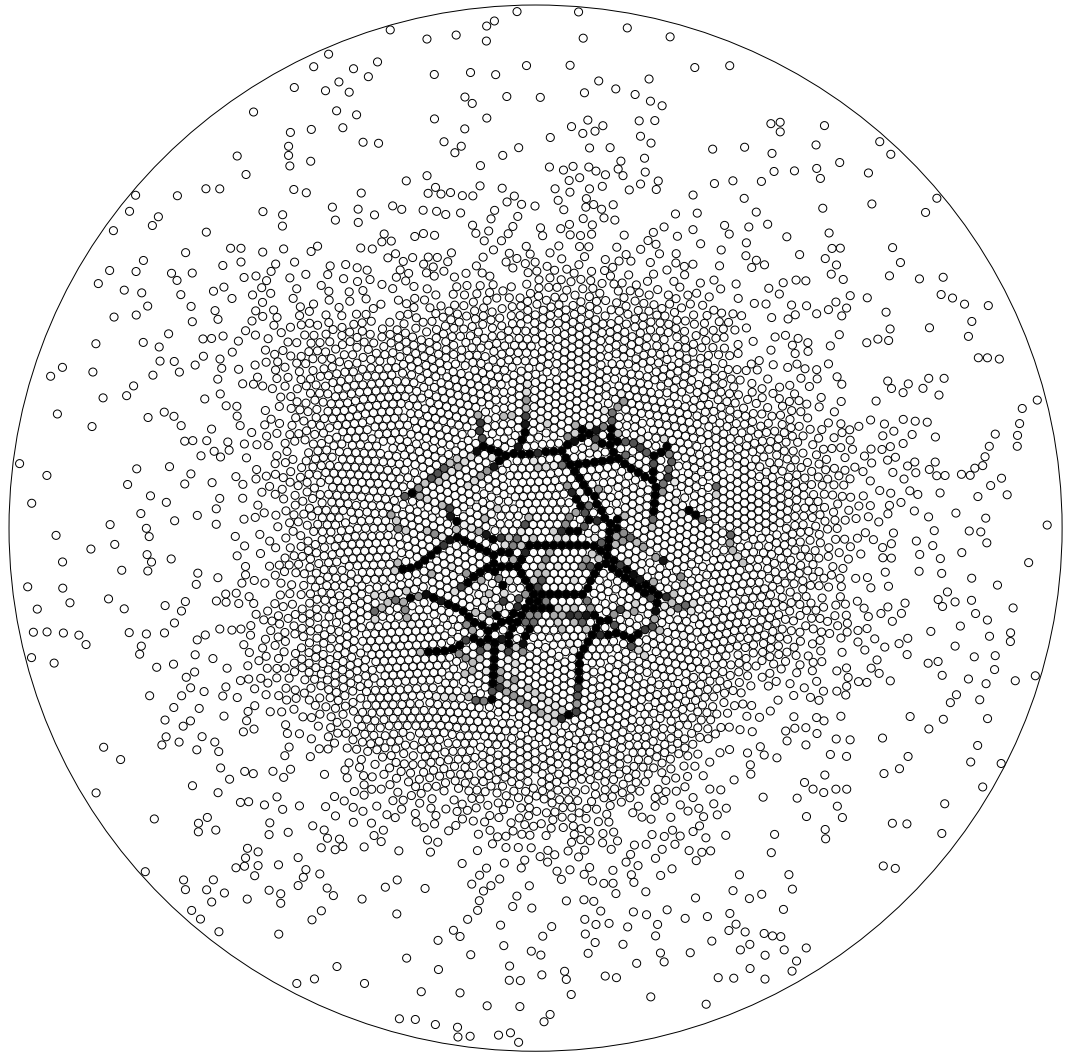,width=14cm,angle=0}}
  \caption{A snapshot from an event-driven Molecular Dynamics 
    simulation of 5000 hard spheres. The wall is kept at a fixed
    temperature. The gray scale codes the relative number of
    collisions experienced by particles during the previous $10^5$
    time steps. One distinguishes three regions: a gas-like state with
    low density, and two high density regions, with and without
    collision chains (c.f.~Fig.~\ref{fig1}).}
  \label{fig:MD}
\end{figure}

The phase of granular system depends on inelasticity of collisions,
$r$, (restitution coefficient
approximation,~\cite{Savage,Haff,Zanetti}), particle density, $\rho$,
particle size, $a$, system size, $L$, and the observation time, $t$.
The external influence of shaking, gravity, boundaries enters through
the above parameters. Granular temperature, or any other
characteristic of the rate of bulk-averaged motion doesn't appear on
the phase diagram since there is no characteristic energy scale for
hard-core interaction assumed here. As we shall see below, a
combination, $\rho{L}a^{d-1}$, where $d$ is the system dimension, is
playing the key role.  It represents the average number of particles
inside an imaginary tube of length $L$ and cross-section $a^{d-1}$.
Presently we think that the phase diagram consists of at least three
regions, Fig.~\ref{fig1}. In the region $1-r \ll
(\rho{L}a^{d-1})^{-2}$ the system is gas-like. In the region
$(\rho{L}a^{d-1})^{-2} \ll 1-r \ll (\rho{L}a^{d-1})^{-1}$ the system
is condensed but doesn't collapse, and for $(\rho{L}a^{d-1})^{-1} \ll
1-r$ the system contains inelastically collapsed chains of particles,
mixed with non-collapsing regions (see ~\cite{Young} for the
description of collapse). Fig.~\ref{fig:MD} is a snapshot from an
two-dimensional event-driven Molecular Dynamics (MD) with 5000
particles in a circle the wall of which is maintained at a constant
temperature.  Particles undergo inelastic collisions with a constant
coefficient of restitution (see Eqs.~(\ref{xxx},\ref{xxxx}) below).
When a particle hits the wall it experiences diffusive angular
scattering, while the distribution of the scattered velocity amplitude
is Maxwellian, with the temperature being that of the wall.  After a
period of equilibration the system finds a state drawn in
Fig.~\ref{fig:MD}, which is in many respects a steady state.  In
Fig.~\ref{fig:MD} the gray scale linearly codes the relative number of
collisions experienced by each particle during the previous $10^5$
time steps, where black means high collision rate. One can clearly
distinguish three different regions: (i) close to the wall there is a
gas-like phase with low density where the mean free path is large.
(ii) The region between the center of the bulk and the wall consists
of closely packed particles. We consider this region as condensed
phase. It is well separated from the gas like phase. (iii) Close to
the center of the container we find a region which isn't
distinguishable from the second region by ``naked eye''.  With the
help of gray scale coding one observes collision chains, i.e. almost
linearly arranged chains of particles which made major contribution to
the previous $10^5$ collisions, in comparison to their neighbors.
Quantitatively, about 5 percent of the particles participated in about
96 percent of the collisions.  Sometimes we observed that very few
($\approx 5\dots 10$) particles take almost all of the collisions in a
certain time interval (not shown).  The lengths of the appearing
chains as well as their life times vary irregularly, their statistics
will be discussed in detail elsewhere~\cite{II}. This figures serves
as an illustration that in MD simulation one encounters all three
regions of the phase diagram given in Fig.~\ref{fig1}.

The role of observation time is not included here, and will be
discussed separately below.  Notwithstanding this classification, the
condensed phases may be ordered in space (a crystal) or disordered (a
glass). This different structure-based classification exists in the
elastic limit.  We now present the arguments used for constructing
Fig.~\ref{fig1} beginning from the gas-like phase.

As the parameter~$\rho{L}a^{d-1}$~increases the first condensation or
clustering transition of granular gas occurs.  The above given
criterion has been obtained by Goldhirsch and Zanetti~\cite{Zanetti}
with the usage of granular hydrodynamics~\cite{Savage,Haff}.~This
description is based on the assumption of molecular
chaos~\cite{Bibleten} and Maxwellian distribution functions for
particle velocities~\cite{Savage} which have been questioned and
related to the concept of inelastic collapse~\cite{Kadanoff}. In this
paper we present the limits of validity of granular hydrodynamics, and
then use the hydrodynamical clustering instability to argue that it
occurs before the condition of inelastic collapse is satisfied.  The
latter is the line sep\-arating collapsing and non-collapsing
condensed phases,
$1-r\sim(\rho{L}a^{d-1})^{-1}$,~\cite{Young,Kadanoff}.
 
Inelastic collapse, when a chain of particles experiences an infinite
number of collisions in finite time as discussed by McNamara and Young
~\cite{Young}, is realized inside dense clusters, specified by
granular hydrodynamics. Inside such condensed regions we have
numerically observed {\it collision} chains.  The relation of
hydrodynamic instabilities of collision chains to their collapse and
the importance of the upper line in Fig.~\ref{fig1} will be published
elsewhere ~\cite{II}.

We begin with diluted phase.  The study of granular gases is greatly
simplified by the fact that the {\it binary} collisions dominate and
goes back to the works of Boltzmann and
others~\cite{Boltzmann,Bibleten}. It is unclear {\it a priori},
whether inelasticity represents a regular or singular perturbation.
The first part of this problem is the reduction of Boltzmann equation
to hydrodynamics, it can be studied independently from the known
complications associated with hydrodynamical description, i.e.
divergence of transport coefficients at high orders in spatial
inhomogeneities and in low dimensions~\cite{Mazenko}. The situation is
somewhat analogous to kinetics of phonons where hydrodynamical
reduction and second sound have a restricted range of applicability as
compared to the kinetic equation ~\cite{Gurevich}. More difficult
problem is the validity of the Boltzmann equation itself, or the
hypothesis of molecular chaos.

\noindent The Boltzmann equation for inelastically colliding identical particles
reads~\cite{Boltzmann,Bibleten,Isihara}\vs
\begin{equation}
\label{boltzmanneq}
\partial_t f + \vec{v} \partial_{\vec{x}} f = \hat{C}(f,f)
\end{equation}
where $f(\vec{v}, \vec{x}, t)$ is the velocity distribution function,
and $\hat{C}$ is the bilinear collision operator\vs\vs\vs
\begin{equation}
\label{collisionoper}
\hat{C}(f,f) = \int \left\{\frac{1}{r^2}f^\prime_1 f^\prime - 
f f_1\right\} \left|\vec{v} - \vec{v}_1\right| d\sigma d\vec{v}_1~.
\end{equation}
The differential cross-section of two spheres of radius $a$ is
$d\sigma = \pi a^2 \sin^2\theta d\theta\,/\,2$, where $\theta$ is
the angle between the vectors $\vec{q}$ and $\vec{v} - \vec{v}_1$. An
incoming collision event to the state $\{\vec{v}, \vec{v}_1\}$ occurs
between particles with velocities $\vec{v}^{~\prime}$ and
$\vec{v}^{~\prime}_1$,
\begin{eqnarray}
  \vec{v}^{\,\prime} &=& \vec{v} - \frac{1+r}{2r} \vec{q}\left[\vec{q}
    \left(\vec{v} - \vec{v}_1\right)\right]\label{xxx}\\ 
  \vec{v}^{\,\prime}_1 &=&
  \vec{v}_1 + \frac{1+r}{2r} \vec{q} \left[\vec{q} \left(\vec{v} -
      \vec{v}_1\right)\right]~,\label{xxxx}
\end{eqnarray}\vs
\noindent 
here $\vec{q}$ is a unit vector pointing from the center of the sphere
1 to the center of the sphere 2, and $0 < r < 1$ is the so-called
coefficient of restitution, it models energy losses in the center of
mass reference frame~\cite{Young,McNamara}. In the case $r = 1$ one
recovers the usual elastic limit.

We now consider a uniform cooling of granular gas without gravity
inside an elastic two-dimensional circle when inelasticity is too
small for developing clustering instability~\cite{Zanetti}. The
problem is (hopefully!) isotropic and homogeneous and the distribution
function, $f_0$, must depend only on the absolute value of particle
velocity, $v$, i.e., on kinetic energy, $E = mv^2/2$. When the initial
distribution is forgotten one may search for a similarity solution
\begin{equation} 
\label{similarity}
f_0(E,t) = \frac{n}{T(t)} \phi\left(\frac{E}{T(t)}\right)~,
\end{equation}
where $T(t)$ is a single scale of the kinetic energy. The
normalization condition to the number density of particles $n$ is
$\int dz \phi(z) = 1$. Eqs.~(\ref{boltzmanneq}, \ref{collisionoper})
result in an integro-differential equation for function $\phi(z)$
\begin{eqnarray} 
\label{simeq}
&&- A(r) \left( \frac{d}{2}\phi + z \frac{d\phi}{dz} \right) = \sqrt{2}
\int d\alpha dz_1 d\sigma g\left(z_1\right)\\
&&~~~~~~~~~~~\times\left\{ \frac{1}{r^2}
  \phi^\prime_1 \phi^\prime - \phi\phi_1\right\} \left(z - z_1 -
  \sqrt{zz_1} \cos\alpha\right)^{1/2}\nonumber~,
\end{eqnarray}
and in an equation for $T(t)$,\vs 
\begin{equation}
\label{hydone}
dT/dt = - A(r) T^{3/2}~.
\end{equation}\vs 
Here $A(r)$ is the constant introduced by separation of variables,
$\alpha$ is the angle between vectors $\vec{v}$ and $\vec{v}_1$, and
$g(z)$ is the density of states. Examination of the collision integral
allows  to get asymptotic form of $\phi(z)$ at large $z$.  Namely, it
can be shown that the incoming term contains extra factor
$O\left(z^{-1/2}\right)$~ [this is contributed by the events
$\left(\vec{q}\,\vec{v}\,\right) \propto z^{-1/2} $~] and is small as
compared to the outgoing term. The asymptotic form of
Eq.~(\ref{simeq}) is $-A(r) z (d\phi/dz) \sim \phi z^{1/2}$, up to a
number. Therefore,\vs
\begin{equation}
\label{ansfirst}
\ln \phi (z) \sim \sqrt{z}/A(r)~.
\end{equation}\vs 
The moments of this function converge, and one can restrict oneself to
the hydrodynamical description (\ref{hydone}) of the granular
temperature which gives
\begin{equation}
\label{tempone}
T(t) = T_0 \left[ 1 + \frac{t}{2}A(r)\sqrt{T_0}\right]^{-2},
\end{equation}
and $T(t) \propto t^{-2}$ at large $t$, ~\cite{Haff}. One can get this
power law from dimensional arguments. $A(r)$ remains undetermined, it
can only be found from the full solution of Eq.~(\ref{simeq}).
Asymptotically, $A(r) \propto (1-r)$ when $r$ is close to 1. The
enhanced population of large energies, (\ref{ansfirst}), dominates at
$z \gg 1/A^2(r)$. For lower energies the distribution is Maxwellian.

\begin{figure}[p]
  \centerline{\psfig{figure=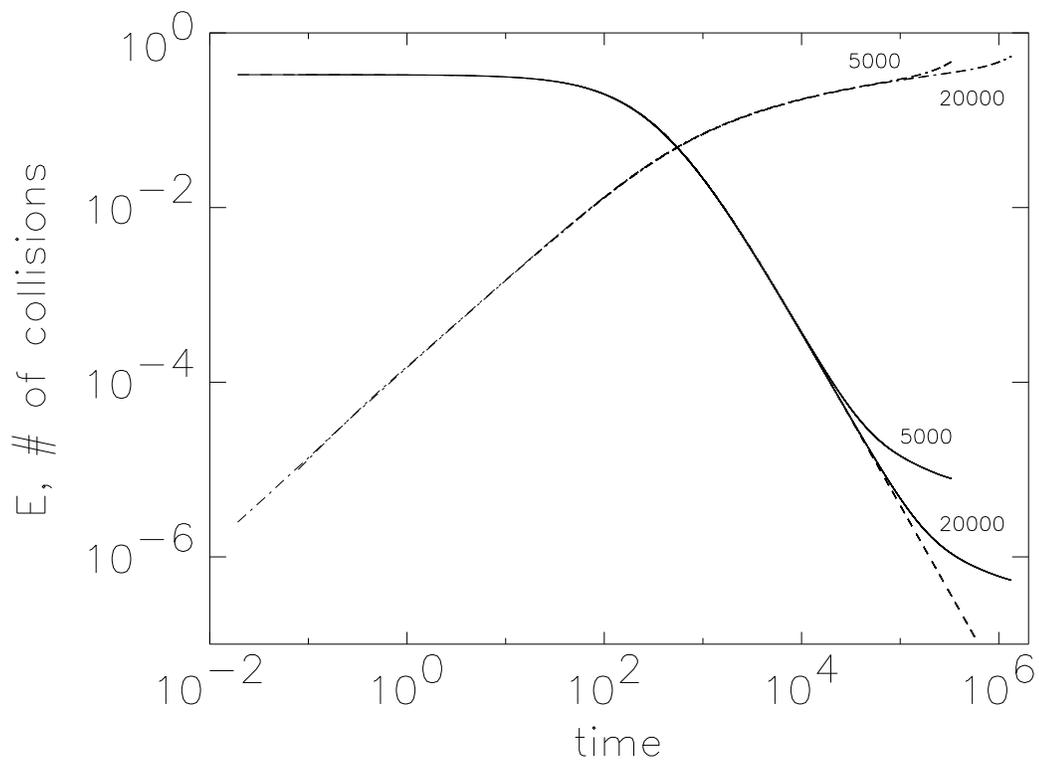,width=14cm,angle=0}} \vskip 2 cm
\caption{The kinetic energy $E$ per particle over time $t$ from MD 
  simulations. Energy decay curves for systems with 5000 and 20000
  particles are identical for a very long time (solid lines) The
  dashed line shows the prediction of Eq.~[8]
  with $T_0 = 0.333$ and $A=0.01$. The dash-dotted lines display the
  cumulative number of collisions per particle scaled by factors
  $10^{-8}$ (5000) and $4 \times 10^{-8}$ (20000).}
\label{fig2}
\end{figure}

\begin{figure}[p]
\centerline{\psfig{figure=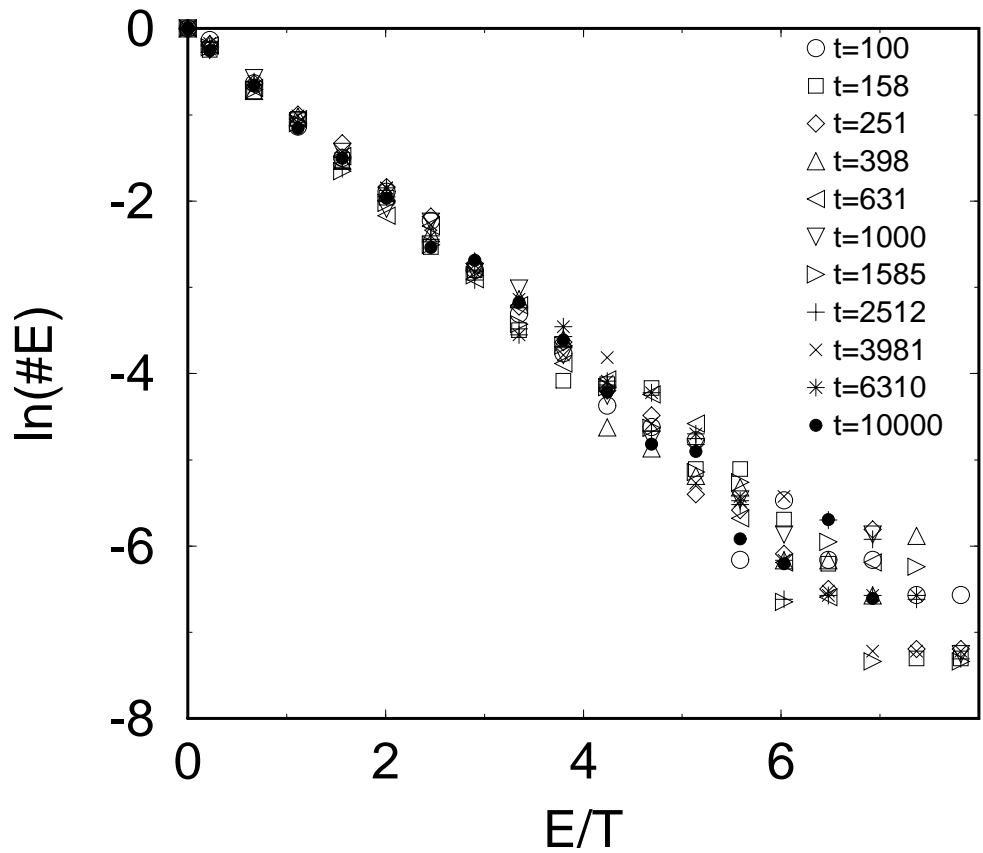,width=14cm,angle=270}}
\vskip 1cm
\caption{Evolution of the scaled energy distribution function in 
 MD simulations. Different symbols used for different times as 
 shown by insets. Statistically, the distribution is 
 indistinguishable from Maxwellian within the given range.}
\label{fig3}
\end{figure}

Two sets of MD results by an event-driven code are obtained with 5000
(20000) particles of unit radius in a circle with radius 130 (260). In
both cases the surface fraction covered by particles is 50/169.  Units
of mass, length and time are arbitrary, initial velocity distribution
is taken to be uniform in velocity in the square $-1 \le v_x, v_y \le
1$.  The employed model of collision is the same as given above, $r =
0.999$.  Fig.~\ref{fig2} shows the temporal evolution of mean kinetic
energy and the number of collisions occurred. The fit to the energy
decay over the entire range is achieved with the help of
Eq.~(\ref{tempone}).  The distribution function is self-similar over
four orders of energy decay as is seen in Figs.~\ref{fig2},
\ref{fig3}. The latter figure shows the energy distribution function
with the argument rescaled by $\langle E(t)\rangle = T$ for different
times.  One can see that the hydrodynamical description is quite
precise.  The system remains uniform and gas-like.

However, at times~$t \sim 10^4 - 10^5$~we observed appearance of a
rarefied space in the center of the circle followed by
``condensation'' of granular gas on the wall. This is a novel type of
transition, which is different from clustering~\cite{Zanetti} or
collapse~\cite{Young}.  One can understand the growth of fluctuations
using the following argument.  The mean free path $l$ is given by
$l=1/n\sigma = a/c$ with the volume fraction $c=na^d$ in a $d$
dimensional system with concentration $n$. The diffusivity of
particles decreases in time like $D(t) \sim (\delta{v})l \sim
l\sqrt{T}$, where $T$ is given in energy units. At large $t$
(Eq.~(\ref{tempone})),
\begin{equation}
T=\frac{4}{\left(tA\right)^2} \sim \left[\frac{l}{t(1-r)}\right]^2~.  
\end{equation}
Therefore,
\begin{equation}
D\sim \frac{l^2}{t(1-r)} = \frac{\left(a/c\right)^2}{t(1-r)}~,   
\end{equation}
and the diffusion length of particles in time $t$ is
\begin{equation}
l_D \sim \sqrt{\int{D}(t)\,dt} \sim \frac{\left(a/c\right)^2}{1-r}
\,\ln^{1/2}\left(\frac{t\,v_0}{a}\right)~,   
\end{equation}
where $v_0$ is the typical initial velocity. The number of particles
inside the sphere of radius $l_D$ is $N\sim n\,l_D^d =
c\left(l_D\,/\,a\right)^d$. These particles preserve a fluctuational
collective velocity $v_0\sqrt{c\left(l_D\,/\,a\right)^d}$. Equating
this velocity to thermal velocity $\delta\,v \sim \sqrt{T(t)} \sim
\left(a/c\right)\,/\,\left(t\,(1-r)\right)$ yields the time when the
collective velocity exceeds the fluctuational individual velocity
$\delta\,v$
\begin{equation}
\label{chaos}
t_c \sim \frac{a}{v_0} 
\frac{c^{\frac{d-3}{2}}}{(1-r)^{\frac{4+d}{4}}} \ln^{-d/4}
\left(\frac{c}{1-r}\right),
\end{equation} 
given with logarithmic precision. Granular flow becomes
fluctuationally {\it supersonic} at $t \gg t_c$ and molecular chaos
assumption along with the Boltzmann equation and granular
hydrodynamics are no longer applicable.  The r.h.s. of Eq(\ref{chaos})
can be evaluated, and the corresponding time is $5.0\times 10^4$.
Given that the numerical prefactor in the estimate (\ref{chaos}) isn't
known, one finds a very satisfactory agreement between the estimate
(\ref{chaos}) and the time where hydrodynamical prediction deviates
from the MD time-trace of kinetic energy density in Fig.~\ref{fig2}.
At time $t_c$ groups of particles of the size $l_c \sim [\int^{t_c}dt
D(t)]^{1/2}$ become special. Their mutual encounters lead to
structures containing travelling {\it shock waves}.  The difference
between the systems with 5000 and 20000 particles as seen in
Fig.~\ref{fig2} is due to the fact that the regions of the size $l_c$
occupy a different fraction of the area, and in the second case it
takes longer for inhomogeneities to evolve from the size $l_c$ up to
the system size.  The observed fluctuational transition is the reason
why observation time may explicitly appear in the phase diagram.

We now add a constant energy supply from the border, and allow the
system to reach a steady state. The system is no longer uniform in
space, and we assume that the corresponding length scales exceed the
mean free path, $l$, everywhere.  The similarity Ansatz
Eq.~(\ref{similarity}) with a non-uniform temperature $T(x)$ can be
used again. A study of the asymptotic form of the collision term shows
that in the regions of the phase space where the detailed balance is
absent the incoming term is again small in $O(z^{-1/2})$ times as
compared to the outgoing term, and no global balance is possible. If
$1-r \ll 1$, the detailed balance approximately holds for the
energies, $z \ll (1-r)^{-1}$, above this range there are effectively
no particles. Therefore, we find that the distribution function is
almost Maxwellian for the energies less than $T/(1-r)$ and small
otherwise,\vs
\begin{equation} 
\label{alamaxwell}
\phi(z) = \left\{ \begin{array}{rl} 
\exp(-z) ~~~ & ~~~ z \ll (1-r)^{-1} \\
 0 ~~~ &~~~z \gg (1-r)^{-1}~.
\end{array}\right.
\end{equation}
All the moments converge, and one can justify the hydrodynamical
reduction, which reads~\cite{Savage,Zanetti,Kadanoff}\vs
\begin{eqnarray}
&&\nabla P(\rho,T) = 0\\
&&\nabla q - B(r) T /\tau_c(\rho,T) = 0~,
\end{eqnarray}
\label{hydrotwo}
\noindent
where $P$ is the granular pressure, and $q$ is the thermal flux
~\cite{Savage,Jackson}. This system can be analyzed.  To understand
what happens as one increases inelasticity (or the number of
particles) it is useful to consider the dilute limit approximation
when $q = \lambda(T)\nabla{T}$, $P = \rho{T}$, where $\lambda(T) =
\gamma_1 T^{1/2}$ is the thermal conductivity, $B(r) \propto (1-r)$ is
the energy fraction lost per collision. $\tau_c^{-1}(n,T) =
\gamma_2\rho T^{1/2}$ is the time between collisions. We found that
the hydrodynamical solution exists only below a threshold,
\begin{equation}
\label{old}
a^{d-1} B^{1/2}(r) \int_0^L \rho ~dx < \xi_d (\gamma_1/\gamma_2) \ln(L/a),
\end{equation}
$\xi_d$ is a pure number. Our hydrodynamical analysis is valid when
~(\ref{old}) is fulfilled; in the opposite case a particle condensate
appears, and system has more than one phase. The left-hand side of
~(\ref{old}) is the generalization of the parameter $\rho{L}a^{d-1}$
for non-uniform case.  Its region of applicability is more general
than the derivation given above.  Eq(\ref{old}) describes with
logarithmic precision the same clustering transition as discussed by
Goldhirsch and Zanetti ~\cite{Zanetti}.

Returning to the question of validity of hydrodynamical reduction of
the Boltzmann equation we note that the complete system of equations
for granular hydrodynamics ~\cite{Mazur,Savage,Jackson} for density,
linear and angular velocity density and granular temperature(s) has a
shortcoming.  It follows from Eqs.~(\ref{ansfirst},\ref{alamaxwell})
that in sufficiently inelastic complex flows where changes occur in
space and time the granular temperature cannot be introduced. Indeed,
if we restrict ourselves with the first moment of the energy
distribution in situations when this distribution changes its
functional form, we wouldn't be able to close the hydrodynamical
reduction. Therefore, in general, the full system of hydrodynamical
equations~\cite{Mazur,Savage} cannot serve as a quantitative
description. The situation is a bit easier, though, than in the
kinetics of phonons: only kinetic coefficients are not exact, and the
uncertainty is in numerical prefactors, which depend on local
distribution function. To give an example about difficulties with
phonons it is sufficient to recall that the heat transfer may be
nonlocal~\cite{Levinson}.

Sufficiently inelastic problems require a mixed kinetic-hydrodynamical
description. In this case, similar to kinetics of
phonons~\cite{Levinson}, one finds a reduced kinetic equation for the
isotropic part, $f_0$, of the distribution function
\begin{equation}
\label{mixed}
\partial_t f_0 - \nabla \vec{J}\left[f_0\right] = 
\hat{C}\left(f_0,f_0\right),
\end{equation}
where diffusion-like flux $\vec{J}$ is defined as
\begin{equation}
\label{diffoperator}
\vec{J}(f) = \frac{4E}{9m}\left[\frac{\delta\hat{C}}{\delta f}\right]^{-1}
 \nabla f.
\end{equation}
The inverse integral operator $\left[\frac{\delta\hat{C}}{\delta
    f}\right]^{-1}$ is fixed by particle number constraint, i.e. the
result of applying this operator to $f_0$ must have vanishing norm.
Similar expressions arise for all dissipative coefficients.  Despite
the cumbersome appearance, Eq.~(\ref{mixed}) together with the
remaining hydrodynamic equations for density and momenta conservations
offers a reduction from nine to five-dimensional space of arguments.

Our study of inelastic gas with binary collisions provided an
opportunity to justify granular hydrodynamics for simple flows and for
all flows in dilute and sufficiently elastic systems.  This allows to
discriminate between clustering instability discussed by Goldhirsch
and Zanetti ~\cite{Zanetti} and inelastic collapse ~\cite{Young}.  In
complex inelastic flows the predictions of granular hydrodynamics are
valid on the order of magnitude almost everywhere, and such precision
is comparable to that in Fig.~\ref{fig1}.  On the basis of this study
we constructed the granular phase diagram and identified a novel type
of supersonic fluctuational phase transition, leading to shock waves,
which is different from clustering and collapse.

We acknowledge constant attention by H.~Jaeger and discussions with
other members of the Chicago Sand Club, Y.~Du, E.~Grossman,
L.~Kadanoff, S.~Nagel, C.~Salue\~na, and T.~Zhou.  We thank P.
Chaikin, R. Jackson, R.~Lieske, L.~Schimansky-Geier and F.~Spahn for
useful discussions. S.~E.~E. is indebted to D.~Frenkel for saving the
book~\cite{Boltzmann} from being trashed in the Physics Library of the
University of Illinois at Urbana-Champaign. This work was supported in
part by the MRSEC Program of the National Science Foundation under the
Grant Number DMR-9400379.







\end{document}